\documentclass[12pt,a4paper]{conference}

\usepackage{fancyhdr}
\usepackage{graphicx,amsmath,amssymb,cite}
\usepackage{multind}

\begin{document}

\Chapter{HOW IS EXOTICS PRODUCED ? \\ WHERE TO SEARCH FOR IT ?}
           {How is Exotics Produced?}{Ya. Azimov \it{et al.}}
\vspace{-6 cm}\includegraphics[width=6 cm]{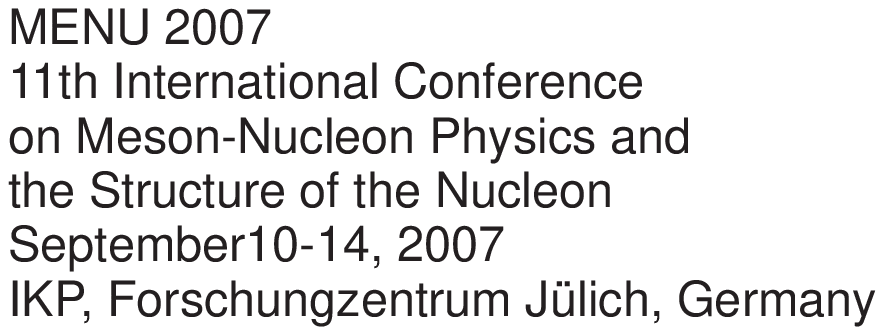}
\vspace{4 cm}

\addcontentsline{toc}{chapter}{{\it Ya. Azimov}}
\label{authorStart}

\begin{raggedright}


Ya.Azimov$^{\star}$$^,$\footnote{E-mail address:
azimov@thd.pnpi.spb.ru}~, K.Goeke$^{\%}$,
and I.Strakovsky$^{\#}$
\bigskip\bigskip


$^{\star}$Petersburg Nuclear Physics Institute,
Gatchina, Russia\\ $^{\%}$Ruhr-Universit\"at, Bochum,
Germany\\ $^{\#}$George Washington University, Washington, D.C.,
USA\\

\end{raggedright}


\begin{center}
\textbf{Abstract}
\end{center}
On the basis of existing data, we suggest such a mechanism of
production for exotic hadrons that can explain, at least
qualitatively, why the $\Theta^+$-baryon is seen in some
experiments and not in others. With our hypothesis, production
of exotic hadrons is a new kind of hard processes. We also can
propose new experiments to check (and confirm?) existence of
exotics and to provide new important information about both
exotic and conventional hadrons\footnote{The talk is based on
the papers~\cite{ags}.}.

\newpage
\section{Introduction}

The problem of exotic hadrons (\textit{i.e.}, non-$qqq$
baryons and/or non-$q\bar{q}$ mesons) stays open. Theoretical
studies are uncertain and do not provide any convincing
explanation, whether and why such hadrons might/should not
exist. But experimental situation is also uncertain in respect
to their existence~\cite{PDG}.

There are mesons (scalar mesons and some of recently discovered
ones) interpreted in the literature as tetraquark ($2q2\bar{q}$)
states, but this interpretation is strongly model dependent.
Their conventional $q\bar{q}$ interpretation cannot be rejected
as well, while rare experimental evidences for explicitly
non-$q\bar{q}$ mesons are not sufficiently reliable yet.

The situation may be different for baryons. There are
experimental evidences for three explicitly exotic states:
$\Theta^+(1530)$, $\Theta_c^0(3100)$, and $\Xi_{3/2}^{--}(1860)$
(or $\Phi^{--}(1860)$)~\cite{PDG}.  However, each of the two
latter states was seen by one group only. They have not been
found in later dedicated experiments, and we will not discuss
them here.

At present, more crucial looks the existence or non-existence
of the $\Theta^+$. The corresponding information is much more
copious than for any other exotic hadron candidate. But the
problem is that there are both positive and negative results,
of several groups on each side.

With such data, one can take a viewpoint that all positive
results might emerge as statistical fluctuations and not reveal
a true physical object. It would be strange, however, to have
the same fluctuation in data of more than ten independent groups
studying very different processes. Moreover, in such a case we
should still live with the open question of what prevents exotic
hadrons from being existent and observed.

An opposite viewpoint is that the $\Theta^+$, as a
representative of exotic hadrons, does exist and has properties
corresponding to the published positive evidences: rather low
mass and unexpectedly narrow width. Then the problem is whether
such an unfamiliar production mechanism may exist, with which
all the present positive and null data are consistent to each
other.

\section{$\Theta^+$ Production Mechanism}

The ZEUS Collaboration was the first experimental group to study
not only existence of the $\Theta^+$, but also its production
properties in Deep Inelastic Scattering (DIS). They compared,
in the same kinematical region, characteristics of the three
baryons, $\Lambda(1520),~\Lambda_c^+(2286),$~and
$\Theta^+(1530)$, which could be kinematically similar. However,
all three appeared dynamically different~\cite{zeus1}.

To understand the results, let us first recall the nature of the
DIS process. The target hadron (proton at HERA) looks in this
process as a set of many partons. At the hard stage of the
process, one (or a few number) of the partons is knocked out by
the virtual $\gamma/Z$. After that, the knocked-out parton(s)
and the remnant part of the target hadronize, softly and nearly
independently of each other.

The ZEUS results~\cite{zeus1} show, that production of the
$\Lambda(1520)$ is well described by hadronization of the
knocked-out parton-quark, exactly as was expected. Production
of the $\Lambda_c^+(2286)$ (or its antiparticle) goes in a
different, but also expected way: the virtual $\gamma/Z$
collides with the parton-gluon, they produce the
$c\bar{c}$-pair, which then hadronizes. Contrary to those,
production properties of the $\Theta^+(1530)$ give evidence,
quite unexpectedly, that it comes from hadronization of the
proton remnant.

Since the remnant is, evidently, a many-parton state, we can
generalize this fact as {\bf the Hypothesis}:
\begin{itemize}
\item
Multiquark (exotic) hadrons are mainly produced through
many-parton configurations, which may emerge either as hadron
remnants in hard processes or just as virtual short-term
fluctuations of the initial hadron(s).
\end{itemize}

Note that in terms of Quantum Chromodynamics (QCD) any hadron
may be described by a Fock column, with different components
having different number of partons. In the space-time picture,
the short-term fluctuations of a hadron are related with higher
Fock components. Hadron picture, as seen in DIS, is also related
with higher Fock components. Thus, in framework of our
hypothesis, the $\Theta^+$-production is always due to
short-term fluctuations and, at least some stage of this
process, should have small characteristic time. This means that
the exotics production is, intrinsically, a new kind of hard
processes.  Of course, it differs from DIS and many other hard
processes, that have a continuous parameter to measure the
hardness (the photon virtuality $Q^2$ for DIS). But it is
similar to the heavy quark production, having the quark mass as
a fixed hardness parameter.  For the exotics production,
hardness may be related to the fixed minimal number of
additional $q\bar{q}$ pairs.

\section{Checks for the Production Mechanism}

Now we can suggest some immediate checks for the hypothesis.

The more is virtuality $Q^2$ in DIS, the higher is effective
multiplicity of partons in the target. Therefore, we expect that
increasing $Q^2$ should provide some (logarithmically
increasing?) enhancement of exotics (say, the $\Theta^+$)
production in respect to conventional hadrons. Such expectation
does not contradict to the preliminary ZEUS data~\cite{zeus2},
though present rather large experimental errors do not allow to
make a clear conclusion. The situation reminds the case of the
Bjorken scaling, which looked exact in early data, while later
more exact measurements revealed its violation.

Our hypothesis suggests interesting expectations not only for
DIS, but also for exotics production in ``soft processes''.
If it needs indeed participation of many-parton fluctuations,
then the accompanying hadron multiplicity should be higher than
in conventional hadron production. Because of kinematical
reasons, this should generate energy spectra, which is softer
for exotics production than for production of only conventional
hadrons. Such expectation appears to be in good correspondence
with the recent result of the SVD Collaboration~\cite{svd}, that
in $NN$ collisions at $E_{\rm lab}=70$~GeV the inclusive
spectrum for $\Theta^+(1530)$ is essentially softer than for
$\Lambda(1520)$.

Additional, indirect support to our hypothesis comes from
calculations of the $\Theta^+$ width~\cite{dpF}. They show
that the extremely low experimental value $\Gamma_{\Theta^+} =
(0.36\pm0.11)$~MeV~\cite{dian} can be described if the decay
$\Theta\to KN$ goes mainly to higher Fock components of the
final nucleon.  Our hypothesis applies similar approach to
production processes as well.

Analysis of Ref.~\cite{ags} shows that the hypothesis
provides also qualitative ways to reconcile current positive and
null experiments (in particular, CLAS and LEPS data~\cite{nak}).
It allows as well to suggest new experiments (or modification of
existing ones) which may confirm and investigate exotic hadrons.

\section*{Acknowledgments}
The work was partly supported by the Russian Grant
RSGSS-1124.2003.2, by the Russian-German Collaboration Treaty
(RFBR, DFG), by the COSY-Project J\"ulich, by Verbundforschung
``Hadronen und Kerne'' of the BMBF, by Transregio/SFB Bonn,
Bochum, Giessen of the DFG, by the U.~S.~Department of Energy
Grant DE--FG02--99ER41110, by the Jefferson Laboratory, and by the
Southeastern Universities Research Association under DOE Contract
DE--AC05--84ER40150.


\end{document}